\documentstyle[multicol,aps,prl,epsf]{revtex}

\begin{document}

\draft

\title{Atomic and Electronic Structures of Unreconstructed Polar
MgO(111) Thin Film on Ag(111)}

\author{Manabu Kiguchi, Shiro Entani, Koichiro Saiki}

\address{Department of Complexity Science $\&$ Engineering, Graduate
School of Frontier Sciences, The University of Tokyo, \\7-3-1 Hongo,
Bunkyo-ku,Tokyo 113-0033, Japan}

\author{Takayuki Goto,  Atsushi Koma}

\address{Department of Chemistry, Graduate School of Science, The 
University of Tokyo, \\7-3-1 Hongo, Bunkyo-ku, Tokyo 113-0033, Japan}

\date{\today}

\maketitle

\begin{abstract}

Atomic and electronic structures of a polar surface of MgO formed on
 Ag(111) was investigated by using reflection high energy electron
 diffraction (RHEED), Auger electron spectroscopy, electron energy loss
 spectroscopy (EELS), and ultraviolet photoemission spectroscopy
 (UPS). A rather flat unreconstructed polar MgO(111) 1$\times$1 surface
 could be grown by alternate adsorption of Mg and O$_{2}$ on
 Ag(111). The stability of the MgO(111) surface was discussed in terms
 of interaction between Ag and Mg atoms at the interface, and charge
 state of the surface atoms. EELS of this surface did not show a band
 gap region, and finite density of states appeared at the Fermi level in
 UPS. These results suggest that a polar MgO(111) surface was not an
 insulating surface but a semiconducting or metallic surface.

\end{abstract}

\medskip


\begin{multicols}{2}
\narrowtext

\section{INTRODUCTION}
\label{sec1}

\begin{figure}
\begin{center}
\leavevmode\epsfysize=30mm \epsfbox{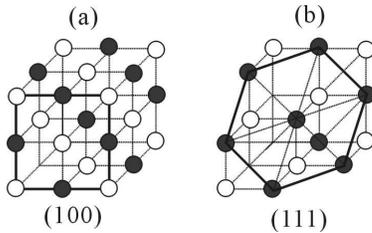}
\caption{Atomic geometry of the non-polar(100) and polar(111) faces of a
 rocksalt crystal.}
\label{fig1}
\end{center}
\end{figure}

\begin{figure}
\begin{center}
\leavevmode\epsfysize=20mm \epsfbox{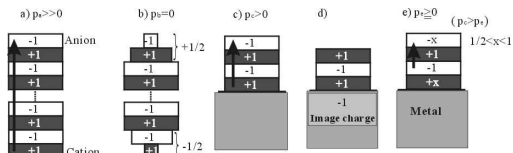}
\caption{Charge distribution and resulting dipole moments (p) for (111)
 surfaces discussed. The numbers in the layers refer to the typical
 charge per atom. (a) Thick (111) film, (b) Reconstructed surface, only
 1/4 and 3/4 of the lattice positions are occupied for the first and the
 second layer, respectively, (c) Neutral 4 ML thick (111) film, and (d)
 3 ML thick (111) film on a metal substrate neutralized by an image
 charge, (e) Neutral 4 ML thick (111) film, the charge of the top and
 bottom atoms reduces to x of the bulk atoms (1/$\leq$x$\leq$1). The
 dipole moment of (e) case is smaller than that of (c) case.}
\label{fig2}
\end{center}
\end{figure}

Polar surfaces have attracted wide attention not only for fundamental
science but also for technological applications, because several
interesting properties, such as novel catalytic activity, two
dimensional electron system, are expected for the polar
surface. Figure~\ref{fig1} shows the atomic geometry of non-polar (100)
and polar (111) faces of rocksalt crystals schematically. The (111)
surface consists of only one atomic species, either cations or anions,
while the (100) surface consists of the same number of cations and
anions. In a rocksalt structure crystal, each atom is surrounded by six
atoms of different species. The coordination number decreases to five
for atoms on a (100) surface, while it changes drastically to three for
atoms on a (111) surface. Therefore, surface energy of the (111) face is
much higher than that of the (100) face, which makes the (111) surface
unstable\cite{1}.  In other words, alternate stacking of cation and
anion layers forms a dipole layer along the [111] direction
(Fig.~\ref{fig2}-a). Accumulation of the dipole layers would produce a
macroscopic electric field, which causes appearance of a flat (111)
surface to be quite unpreferable in the rocksalt structure compounds and
the flat (111) surface does not occur in nature.

From a theoretical viewpoint, the macroscopic electric field in the
polar surface can be cancelled either by surface reconstruction or by
reduction of effective charge of the surface layer. According to Wolf,
the polar (111) surface of rocksalt crystals would be stable under an
octapole termination, which leads to formation of a
2$\times$2 reconstruction on the (111) surface\cite{1}
(Fig.~\ref{fig2}-b). On the other hand, Tsukada and Hoshino pointed out
that the change in charge state of surface atoms could stabilize the
(111) surface of rocksalt structure compounds\cite{2}. If the charge of
top atoms is reduced to half of the bulk atoms, the macroscopic electric
field would not appear.
 
Under these backgrounds, several attempts to grow the polar surface of
rocksalt structure compounds have been performed so far\cite{3}. Metal
oxides have been extensively studied to create crystals or thin films
having (111) surfaces. For a NiO(111) surface, p(2$\times$2)
reconstruction was found, and the surface was determined to be a single
Ni termination with double steps by the measurement of grazing-incidence
X-ray diffraction\cite{4}. The unrecostrcuted surface is stabilized by
some adsorbates. Langell and Berrie reported that the NiO(111) surface
is stabilized by OH adsorbates and that desorption of the hydroxyl
changes the surface to a thermodynamically more stable
NiO(100)\cite{5}. Stabilization by impurities was found also for Pb and
Si on NiO(111)\cite{6}.
  
In contrast with the surface of single crystals, the electrostatic
energy does not reach exorbitant values for the small thickness of a
film (Fig.~\ref{fig2}-c). When the thin film is grown on a metal
substrate, further reduction of energy could come from the image charge
induced in the supporting metal surface. 
Ventrice {\it et al.} have studied
the growth of NiO on Au(111) by LEED and STM, and obtained a 6 ML thick
stable p(2$\times$2) reconstructed NiO(111) film\cite{7}. For a FeO(111)
thin film, interesting results were reported by Koike and
Furukawa\cite{8}. They obtained the FeO (111) p(2$\times$2) surface by
oxidizing Fe(110) and measured the polarization of secondary electrons
emitted from the FeO (111) surface. Ferromagnetic ordering was found for
the FeO (111) p(2$\times$2) surface, although FeO itself was an
antiferromagnetic material with the Neel temperature of 198 K. They
thought that the ferromagnetic ordering comes from the reconstruction of
a polar FeO (111) surface to reduce the large electrostatic surface
energy. On the other hand, the relaxation, that is, the decrease in
interlayer spacing, was observed for the FeO (111) grown on Pt
(111)\cite{9}. In this case, the layer by layer growth was limited to a
maximum of 2.5 ML.

Thus, although the (111) surface of rocksalt structure oxides has been
studied by many groups, there have been few studies on unreconstructed
adsorbate-free (111) surfaces of a rocksalt structure until
now. Furthermore, the electronic structure of the polar surface has not
been clear. In the preset study, we have examined growth of a MgO thin
film on Ag(111) by supplying Mg and O, alternately. MgO has a rocksalt
structure with a lattice constant of 4.21 \AA, while Ag has a fcc
structure with a lattice constant of 4.09 \AA. As the lattice misfit of
MgO to Ag is only -2.9 \%, the first Mg layer is expected to become a
template for the growth of the MgO film along the [111] direction. The
commensurate bonding between Mg and Ag atoms at the interface might help
alternate stacking of Mg and O layers, leading to a flat (111)
surface. Actually, we have observed rather streak RHEED pattern,
implying formation of a flat (111) MgO surface. We have characterized
the grown surface by using EELS and ultraviolet photoemission
spectroscopy (UPS) and discussed stabilization mechanism of the polar
surface.

\section{EXPERIMENTAL}
\label{sec2}

The experiments were performed in a custom-designed ultrahigh-vacuum
(UHV) system with a base pressure of 2$\times$10$^{-8}$ Pa. A
mechanically and electrochemically polished Ag(111) surface was cleaned
by repeated cycles of Ar$^{+}$ sputtering and annealing at 900 K. After
repeated preparation cycles, a sharp RHEED pattern was observed, and no
contamination was detected by AES. A MgO film was grown by alternate
adsorption of Mg and O$_{2}$ on Ag(111) at substrate temperature of 300
K. First, 1 ML (2 \AA) Mg was deposited by evaporating high purity (99.98
\%) Mg onto Ag(111). The growth rate was monitored using a quartz
crystal oscillator. The Mg film was dosed with 10 L O$_{2}$. Real-time
observation of the crystallinity and orientation of films was done by
RHEED. Surface compositions and the electronic structure of the grown
films were investigated {\it in situ} by AES, 
EELS, and UPS. The analyses were
performed with a double pass CMA (PHI 15-255G) in the analysis
chamber. Adoption of a pulse counting detector reduced the probing
electron current down to 1 nA, which reduced surface damage as much as
possible.

\section{RESULTS}
\label{sec3}
\subsection{Heteroepitaxial growth of MgO on Ag(111)}
\label{sec3a}

\begin{figure}
\begin{center}
\leavevmode\epsfysize=50mm \epsfbox{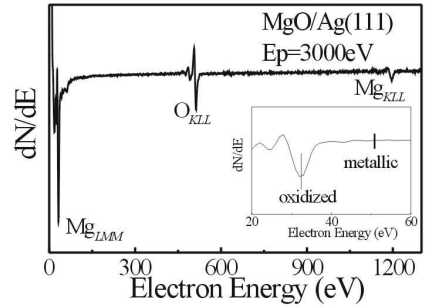}
\caption{Auger electron spectra for the 10 ML thick MgO film on
 Ag(111). The primary-electron energy is 3 keV.}
\label{fig3}
\end{center}
\end{figure}

Figure~\ref{fig3} shows the Auger spectrum of the 10 ML thick MgO film
grown on Ag(111). Disappearance of the Ag LMM Auger peak (351 eV)
indicated that the grown MgO film covered the Ag(111) substrate
completely because the inelastic mean free path of 351 eV electron was
around 1 nm. The chemical state of Mg atoms of the MgO film was known by
the energy of the Mg LMM Auger peak, because the energy of the Mg LMM
Auger peak is 45 eV and 32 eV for metallic Mg and MgO,
respectively\cite{10}. The energy of the Mg LMM Auger peak was 32 eV for
the grown MgO film, showing that Mg was completely oxidized. The
stoichiometry of the MgO film was also examined by the ratio of the Mg
KLL Auger peak intensity to the O KLL Auger peak intensity. Considering
the Auger electron emission probabilities\cite{11}, the ratio of the
amount of Mg to that of O was 1 : 0.92$\pm$0.18\cite{12}, supporting the
idea that a stoichiometric MgO film was grown on Ag(111).

\begin{figure}
\begin{center}
\leavevmode\epsfysize=50mm \epsfbox{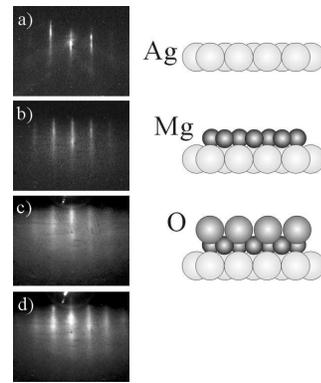}
\caption{A typical sequence of RHEED patterns and schematic models
 during the growth of the MgO film on Ag(111) at a substrate temperature
 of 300 K. a) Ag(111), b) 1 ML Mg/Ag(111), c) 2 ML MgO/Ag(111), d) 10 ML
 MgO/Ag(111). The incident beam was parallel to the [1$\bar{1}$0]
 azimuth of the Ag(111).}
\label{fig4}
\end{center}
\end{figure}

Figure~\ref{fig4} shows the typical sequence of RHEED pattern during the
growth at a substrate temperature of 300 K. The incident electron beam
was parallel to the [1$\bar{1}$0] azimuth of the substrate. The result
of RHEED patterns indicated that the MgO film grew heteroepitaxially on
Ag(111). The epitaxial orientation of the MgO film was determined to be
(111)$_{MgO}$//(111)$_{Ag}$ and 
[1$\bar{1}$0]$_{MgO}$//[1$\bar{1}$0]$_{Ag}$. The half-order streaks did
not appear during the growth, showing that the (1$\times$1)
unreconstructed MgO(111) film was grown on Ag(111). Streaks in RHEED
patterns indicated that a rather flat (111) surface could be
obtained. The RHEED pattern became blurred with increasing film
thickness, suggesting that a thick MgO(111) film was unstable.

The in-plane lattice constant of the MgO(111) film was calculated from
the spacing between streaks in the RHEED pattern. For the 10 ML thick
MgO(111)/Ag(111), the in-plane lattice constant was determined to be
3.28$\pm$0.03 \AA \cite{12}, which was +10 \% larger than that 
of bulk one (2.97 \AA). The in-plane lattice constant
was 3.25$\pm$0.03 \AA {} for the 2
ML thick film and did not change with film thickness, indicating that
the expansion was uniform throughout the epitaxial layer. The increase
of the in-plane lattice constant leads to reduction of the surface
electron density, which might help a decrease in the electrostatic
energy of the MgO(111) film.

\subsection{Stability of the (1$\times$1) MgO(111) film on Ag(111)}
\label{sec3b}

Having established the existence of the (1$\times$1) unreconstructed
MgO(111) film on Ag(111), we should discuss why this structure is stable
in spite of the previously mentioned arguments on the instability of a
polar surface of ionic crystals\cite{1}. We think that the stabilization
of the (1$\times$1) unreconstructed MgO(111) film can be explained in
terms of interaction between Ag and Mg atoms at the interface and charge
state of surface atoms.
 
In the previous study, a single crystalline MgO(100) film grew
heteroepitaxially on Ag(100) by evaporating MgO congruently from an
electron beam evaporator(EB)\cite{13}. On Ag(111), however, a MgO film
could not grow heteroepitaxially by EB. Heteroepitaxial growth of a
single crystalline MgO film was achieved by alternate adsorption of Mg
and O$_{2}$ on Ag(111). These facts indicate that the strong interaction
between Mg and Ag atoms plays a decisive role in stabilizing the
(1$\times$1) MgO(111) film. The free energy of the interface
($E_{inter}$) is smaller for the case with coherent Mg-Ag bonds as
compared with the case, in which fine particles with thermodynamically
more stable \{100\} faces grow on Ag(111) without coherent Mg-Ag
bonds. The structure of the grown film is determined to minimize the sum
of the electrostatic energy of the film ($E_{electro}$) and
$E_{inter}$. Here, we should notice that $E_{electro}$ of a (111) film
does not reach exorbitant values for ultrathin films (see
Fig.~\ref{fig2}-c), because $E_{electro}$ is proportional to the film
thickness. Therefore, the gain in interaction between Mg and Ag
overcomes the disadvantage of the electrostatic energy, and the
(1$\times$1) unreconstructed MgO(111) film is grown on Ag(111). In
addition to the strong interaction with Mg atoms, the Ag substrate plays
another important role for stabilization of the MgO(111) film. The MgO
film was prepared by alternate adsorption of Mg and O$_{2}$. Therefore,
the film is unstable due to breakdown of charge neutrality
(Fig.~\ref{fig2}-d), when the film thickness is odd number (Mg top). On
the metal substrate, however, the image charge is induced in a metal,
helping stabilization of the film with odd layers.
 
Besides the strong interaction between Mg and Ag atoms at the interface,
the stabilization of MgO(111) can be explained in terms of charge
state. Due to the reduction of the coordination number, Madelung
potential largely decreases for the surface O$^{2-}$ on the (111)
surface. The binding energy of the 2p band, thus, decreases for the
surface O$^{2-}$, and the Fermi level might be located in the 2p band. In
such a case, electrons flow out of the upper valence states of the
surface O$^{2-}$, and the charge of the surface O$^{2-}$ is reduced. 
The charge of the Mg$^{2+}$ at the interface would 
also decrease to keep the film neutral
(Fig.~\ref{fig2}-e). The decrease in the charge reduces a macroscopic electric
field, and the instability of the film should be remedied. That is the
second point to stabilize the polar MgO(111) film on Ag(111). Here we
should notice that the reduction in charge of surface O$^{2-}$ leads to
reduction of Madelung potential. Therefore, the amount of the charge
transfer is determined under the condition, in which both Madelung
potential and the charge of surface O$^{2-}$ are self consistent. Tsukada and
Hoshino studied the O-terminated MgO(111) surface by DV-X$\alpha$ calculations,
and revealed that the charge of the top and bottom atoms is reduced to
half of the bulk atoms\cite{2}. When the charge of the surface atom is
reduced to half of the bulk atoms, the macroscopic electric field would
not appear, irrespective of the film thickness. In the present study, a
thick MgO(111) film could not be grown on Ag(111), implying that the
charge of the topmost O atoms did not reduce to half of the bulk one.

Finally, we would discuss the surface reconstruction of the MgO(111)
film. Since the surface energy of a \{100\} face is lower than that of a
\{111\} face in rocksalt structure compounds, \{100\} faces are expected
to appear for growth normal to the (111) face. In case of
NaCl/GaAs(111)A, triangular pyramids with three 
exposed \{100\} faces grow
on the substrate\cite{3,14}. As discussed in Introduction, a
p(2$\times$2) reconstructed NiO(111) film is grown on Au(111)\cite{7},
and the p(2$\times$2) structure corresponds to the smallest triangular
pyramids surrounded by \{100\} faces. Here, we should notice that both
films were grown by congruent evaporation of NiO or NaCl at the
substrate temperatures higher than 420 K. Therefore, the grown film
forms in a thermodynamically stable structure. In the present case,
however, we have grown a metastable flat MgO(111) film by alternate
adsorption of Mg and O$_{2}$ at 300 K. Once a flat (111) film is formed
on Ag(111), activation energy is needed to change the metastable (111)
film into stable triangular pyramids surrounded by \{100\}
faces. Therefore, the flat unreconstructed (111) MgO film remains in a
metastable form on Ag(111).

\subsection{Electronic structure of MgO(111)}
\label{sec3c}

As we could prepare the unreconstructed polar MgO(111) surface, the
electronic structure of this surface was investigated by EELS and UPS,
comparing with that of a non-polar MgO(100) surface\cite{13,15}. Absence
of Ag LMM Auger peak in the 10 ML MgO(111)/Ag(111) assures that EELS and
UPS probe the MgO film with influences of the substrate negligible,
because their probing depth is smaller than or equal to that of AES.

\begin{figure}
\begin{center}
\leavevmode\epsfysize=50mm \epsfbox{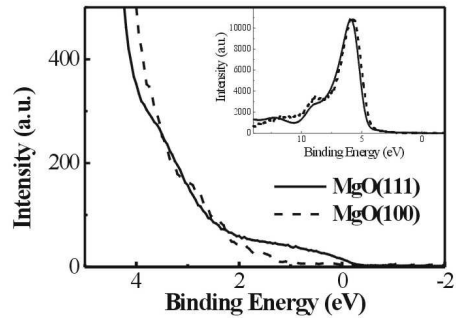}
\caption{UPS of the polar MgO(111) surface taken with He I
 source. Spectra of the MgO(100) surface is shown for comparison.}
\label{fig5}
\end{center}
\end{figure}

Figure~\ref{fig5} shows UPS spectrum for the MgO(111) surface measured
with a He I (21.2 eV) source. For comparison, the spectrum of the 10 ML
thick MgO(100) film on Ag(100) is included in the same figure. The main
features for MgO(111) were almost similar to those for MgO(100) except
around the Fermi level. For MgO(100), there were no density of states
(DOS) near the Fermi level, while an appreciable DOS appeared in the
region from 2 eV to the Fermi level for MgO(111). These finite DOS at
the Fermi level did not originate from metallic Mg, since component of
metallic Mg was not observed in AES. Furthermore, we have revealed that
the finite DOS at the Fermi level disappeared even for the incompletely
oxidized film in the previous study\cite{16}. Therefore, the finite DOS
at Fermi level did not originate from metallic Mg or the incompletely
oxidized Mg, the surface states of MgO(111). The UPS results suggested
that the polar MgO(111) surface was a metallic surface.

\begin{figure}
\begin{center}
\leavevmode\epsfysize=50mm \epsfbox{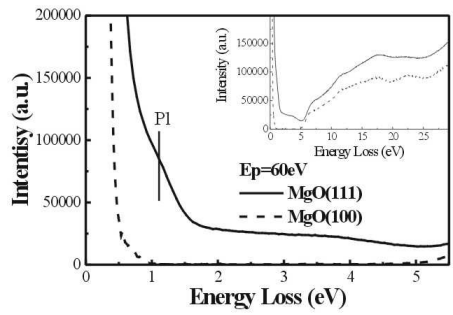}
\caption{EEL spectra of the polar MgO(111) surface (line). The primary
 electron energy was 60 eV. Spectra of MgO(100) surface is shown for
 comparison (dot line).}
\label{fig6}
\end{center}
\end{figure}

Figure~\ref{fig6} shows the EELS of the MgO(111) surface measured with
primary electron energy of 60 eV. For comparison, spectrum of the
MgO(100) surface is also shown in the figure. The structure above 7 eV
for MgO(111) was almost similar to that of MgO(100), showing that a
stoichiometric MgO film grew on Ag(111)\cite{13,15}. On the other hand,
difference appears in the structure below 5 eV between MgO(100) and
(111). In contrast with MgO(100), the clear band gap region was not
observed for MgO(111). In addition, the tail of the elastic peak became
larger and a new peak (p1) appeared at 1.2 eV, indicating existence of
the inelastically scattered electrons with small
loss-energy\cite{17}. Since the AES signal indicative of Mg metal was
not observed at all, the tail or the new peak did not originate from Mg
aggregates in the MgO film.

Due to the low growth temperature and the significant mismatch between
the substrate and the oxide lattice, the film may contain a significant
concentration of defects. Therefore, the spectroscopic features could be
due to emission from defect states in the MgO(111) film. Defective MgO
surfaces have been studied by a variety of techniques such as UPS, EELS,
theoretical calculation, etc. In EELS of defective MgO surfaces, sharp
peaks are observed at 2.3 eV and 5 eV, which are attributed to Mg
vacancy (V center) and O vacancy (F center), respectively. Sharp
structures are also observed for MgO powders by diffuse reflectance
spectra, and these structures at 5.8 and 4.6 eV have been attributed to
ions with four and three-fold ligand coordination\cite{18}. On the other
hand, such sharp peaks were not observed in the EELS of MgO(111) below 5
eV. Recently, we have revealed that metal induced gas states(MIGS) were
formed at the insulator-metal interface\cite{19}. Since the new states
of the MgO(111) film were observed independently of the film thickness,
the states did not originate from MIGS. Therefore, the structure below 5
eV could be assigned to the excitation of electrons in the unfilled band
derived from O 2p, or the excitations of newly appearing electronic
states on the (111) surface.

The UPS and EELS results suggested that the MgO(111) surface was not an
insulating surface but a semiconducting or metallic surface. This
electronic structure of the MgO(111) surface can be explained in terms
of Madelung potential as discussed in the previous section. Compared to
the binding energy of isolated ions, the binding energy of Mg$^{2+}$ 3s
orbital decreases, and the binding energy of O$^{2-}$ 2p orbital
increases by the Madelung potential in a crystal phase. Because of the
decrease in the coordination number, the Madelung potential largely
decreases for the (111) surface, compared with the (100)
surface. Therefore, the band gap is reduced and the MgO(111) surface
changes into a semiconductor or metallic surface. However, a possibility
of certain new electronic state characteristic of the (111) surface
cannot be excluded.
  
\section{CONCLUSIONS}
\label{sec4}
Heteroepitaxial growth of MgO on Ag(111) has been investigated by RHEED,
AES, UPS and EELS. The flat unreconstructed polar MgO(111) surface was
obtained on Ag(111) by alternate adsorption of Mg and O$_{2}$ at
substrate temperature of 300 K. The EELS and UPS of the (111) surface
were different from those of the (100) surface. For MgO(111), a clear
band gap region was not observed in EELS, while finite density of states
appeared in UPS, implying that the MgO(111) surface was a semiconducting
or metallic surface.

\acknowledgments{
The authors are grateful for the financial support of the Grant-in-Aid
from Ministry of Education, Culture, Sports, Science and Technology 
of Japan.}

\end{multicols}

\begin{references}
\bibitem{12} Error is estimated from the assumed precision of the
 technique (atomic ratio determined by AES: 20 \%, lattice constant
 determined by RHEED: 1\%).
\end{references}
\end{document}